\documentstyle[amssymb,preprint,aps]{revtex}

\input{tcilatex}

\begin{document}
\title{Ni impurity induced enhancement of the pseudogap in cuprate high T$%
_{c}$ superconductors}
\author{A.V. Pimenov$^{1}$, A.V. Boris$^{1}$, Li Yu$^{1}$, V. Hinkov$^{1}$,
Th. Wolf$^{2}$, J.L. Tallon$^{3}$, B. Keimer$^{1}$, and C. Bernhard$^{1}$}
\address{1. Max-Planck-Institute for Solid State Research, Heisenbergstrasse%
\\
1, D-70569 Stuttgart, Germany.}
\address{2. Forschungszentrum Karlsruhe, INFP, D-76021 Karlsruhe, Germany.}
\address{3. Mac Diarmid Institute and Industrial Research Ltd., Lower Hutt,
New Zealand.}
\date{\today}
\maketitle

\begin{abstract}
The influence of magnetic Ni and non-magnetic Zn impurities on the normal
state pseudogap (PG) in the c-axis optical conductivity of NdBa$_{2}$\{Cu$%
_{1-y}$(Ni,Zn)$_{y}\}_{3}$O$_{7-\delta }$ crystals was studied by spectral
ellipsometry. We find that these impurities which strongly suppress
superconductivity have a profoundly different impact on the PG. Zn gives
rise to a gradual and inhomogeneous PG suppression while Ni strongly
enhances the PG. Our results challenge theories that relate the PG either to
precursor superconductivity or to other phases with exotic order parameters,
such as flux phase or d-density wave states, that should be suppressed by
potential scattering. The apparent difference between magnetic and
non-magnetic impurities instead points towards an important role of magnetic
correlations in the PG state.
\end{abstract}


The unconventional normal-state (NS) properties of the cuprate high T$_{c}$
superconductors (HTSC) are a key issue in the search for the superconducting
(SC) pairing mechanism. A prominent feature is the so-called pseudogap (PG)
phenomenon which prevails in underdoped samples where it gives rise to a
gradual and incomplete suppression of the low-energy spin and charge
excitations well above the SC transition temperature, T$_{c}$. It was first
discovered by nuclear magnetic resonance (NMR) \cite{Alloul1} and
subsequently identified in specific heat \cite{Loram1}, angle-resolved
photo-emission (ARPES) \cite{Loeser1}, and also in optical measurements \cite%
{Homes1}. In the latter it appears as a gap-like suppression of the
far-infrared (FIR) c-axis conductivity perpendicular to the CuO$_{2}$
planes. The characteristic energy- and temperature (T) scales of the PG
increase with underdoping whereas on the overdoped side they merge with or
fall below the ones of the SC energy gap \cite{Loeser1,Bernhard1}. Numerous
models have been proposed to account for the PG phenomenon. The most
influential ones can be divided into two categories. In the first one, the
PG is related to a precursor SC state where macroscopic phase coherence is
inhibited by thermal phase fluctuations \cite{Emery1} or pair condensation
is delayed to much lower T than pair formation (Bose-Einstein-condensation
scenario) \cite{Alexandrov1}. The second class assumes a distinct kind of
order that may even compete with SC \cite{Tallon1}. Proposals include
conventional and exotic spin- or charge density wave states \cite%
{Chubukov1,Dahm1,Lee1,Chakravarty1}, the stripe scenario \cite{Loew1}, spin
singlets in the spin-charge separation scenario \cite{Anderson1} or lattice
driven instabilities \cite{DiCastro1}.

A major obstacle for the identification of the PG state is that its low T
properties are obscured by the response of the SC condensate. This problem
can be overcome if SC order is suppressed either with a large magnetic field
or by means of impurities within the CuO$_{2}$ planes (Zn or Ni for Cu) that
cause strong pair-breaking. The former approach requires extremely large
fields and is thus not viable in combination with spectroscopic techniques
that access the relevant electronic energy scales. The latter approach has
been limited to samples with rather low impurity content and thus minor T$%
_{c}$ suppression \cite{Uchida1}.

Here we report on the evolution of the PG in the optical c-axis conductivity
of (Sm,Nd)Ba$_{2}\{$Cu$_{1-y}$(Ni,Zn)$_{y}\}_{3}$O$_{7-\delta }$ single
crystals. Up to 9 \% Zn and 17 \% Ni per Cu atom have been incorporated
allowing for a full suppression of SC even at optimum doping. We find that
non-magnetic Zn impurities and magnetic Ni impurities, both of which
strongly suppress SC, have a profoundly different impact on the PG. The
non-magnetic Zn impurities lead to a slow and inhomogeneous PG\ suppression,
whereas the magnetic Ni impurities are beneficial and strongly enhance the
PG phenomenon. Ni substitution restores a normal state PG even in optimally
doped and slightly overdoped samples where the PG is otherwise absent. Our
observations are at variance with theories that relate the PG to a precursor
SC state. Similar arguments apply for the exotic order parameters of the
flux-phase or d-density wave states where potential scattering from
impurities also gives rise to destructive interference effects. Our data
support the point of view that magnetic correlations play a prominent role
in the PG formation.

High quality (Sm,Nd)Ba$_{2}\{$Cu$_{1-y}$(Ni,Zn)$_{y}\}_{3}$O$_{7-d}$ single
crystals were grown with a flux method under reduced oxygen atmosphere to
avoid the substitution of RE ions on the Ba site \cite{WolfRE}. Zn and Ni
contents were determined by energy dispersive x-ray (EDX) analysis. The use
of rare earth ions with a large ionic radius (Nd or Sm) was found to assist
the incorporation of higher amounts of Zn and Ni impurities. The SC
transitions were determined by SQUID magnetometry. The quoted values of T$%
_{c}$ and $\Delta $T$_{c}$ correspond to the midpoint and 10 to 90 \% width
of the diamagnetic transition. The T$_{c}$ suppression closely follows the
reported trend for Zn impurities \cite{TallonZnPRL} while for the
Ni-impurities it is significantly faster than in Y123 \cite{Williams1,Itoh02}%
. Assuming a similar T$_{c}$ suppression rate for Zn and Ni impurities in
the CuO$_{2}$ planes, we deduce that about half the Ni impurities reside in
the planar Cu sites as opposed to the chain site which is favored in Y123 %
\cite{Williams1,Itoh02}. Zn is known to enter exclusively the planar site
even in Y123. By annealing in flowing O$_{2}$ gas and subsequent rapid
quenching into liquid nitrogen we prepared a series of crystals with
different Zn and Ni concentrations but identical oxygen content. It was
shown by thermo-electric power (TEP) \cite{TallonTEP} and NMR \cite%
{Williams1,Itoh02} that Zn substitution leaves the hole doping state, p, of
the CuO$_{2}$ planes unaffected. We performed TEP measurements on our
crystals which establish that the Ni impurities also do not significantly
alter the hole doping state, p.

The ellipsometry experiments were performed at the infrared (IR) beamline at
the ANKA synchrotron at Forschungszentrum Karlrsuhe, Germany and with a
laboratory-based setup \cite{Bernhard2}. Ellipsometry measures directly the
complex dielectric function without a need for Kramers-Kronig analysis. It
is a self-normalizing technique that allows one to determine very accurately
and reproducibly the dielectric function and, in particular, its T-dependent
changes.

In Figs. 1a to 1c we present spectra of the real part of the c-axis
conductivity, $\sigma _{1c}$, below 2500 cm$^{-1}$ (806 cm$^{-1}\triangleq $
100 meV) for Ni substituted NdBa$_{2}$Cu$_{3-y}$Ni$_{y}$O$_{6.8}$ single
crystals with y=0, 0.03, 0.06 and 0.12. The crystals were annealed at 520 $%
{{}^\circ}%
$C in flowing O$_{2}$ and were underdoped with T$_{c}$=65(4) K for y=0, T$%
_{c}$=20(5)\ K for y=0.03, and T$_{c}$%
\mbox{$<$}%
2 K for y=0.06 and 0.12. It is well established that a slightly higher
oxygen content is required for Nd123 as compared with Y123 in order to
obtain the same doping state in the CuO$_{2}$ planes \cite{Wolfnh}. The TEP
data indicate an almost constant hole doping state ranging from p$\approx $%
0.1 at y=0 to p$\approx $0.11 at y=0.17. Figure 1a shows that the
characteristic spectral features of the Ni-free crystal closely resemble
those of similarly underdoped Y123 \cite{Homes1,Bernhard1}. The narrow peaks
correspond to the well known IR-active phonon modes of the 123 structure. A
broader mode near 420 cm$^{-1}$ that develops in the SC state can be
attributed to a transverse Josephson resonance mode \cite{Grueninger1}. Only
the so-called rare-earth mode exhibits a significant red shift from 190 to
at 175 cm$^{-1}$ that is consistent with the mass ratio of Nd and Y. In the
following we are mostly concerned with the underlying broad electronic
response. In particular, we focus on the suppression of the electronic
conductivity below $\sim $1100 cm$^{-1}$ which develops already in the NS. A
corresponding gap feature was previously observed in underdoped Y123 and has
been associated with the PG \cite{Homes1}. It was shown to exhibit a similar
magnitude, spectral shape and doping dependence as the PG that emerges from
other spectroscopic experiments like ARPES or tunneling spectroscopy \cite%
{Bernhard1}. It has a gradual onset at $\omega ^{PG}\approx 2\Delta ^{PG}$
(marked by arrows) that is characteristic of a k-space anisotropy of the
underlying order parameter. Our most important result is that the energy
scale of the PG in the c-axis conductivity becomes strongly enhanced upon Ni
substitution, contrary to the naive expectation that impurities like Ni
should reduce or at least wash out the corresponding gap features. This is
apparent in Figs. 1a to 1c where $\omega ^{PG}$ increases continuously from
1150 cm$^{-1}$ at y=0.0 to about 2050 cm$^{-1}$ at y=0.12. The evolution of $%
\omega _{c}^{PG}$ as a function of Ni content for the entire series is
displayed in\ Fig. 1d. We emphasize that the TEP data establish a nearly
constant hole doping upon Ni substitution, if at all p exhibits a slight
increase. The dramatic Ni induced increase in $\omega ^{PG}$ thus cannot be
explained in terms of a very large Ni-induced hole depletion. We also note
that the corresponding in-plane response remains metallic even at y=0.17,
albeit with a largely enhanced scattering rate (not shown here).

The spectra for a series of correspondingly underdoped NdBa$_{2}$Cu$_{3-y}$Zn%
$_{y}$O$_{6.8}$ crystals with y=0.0 (T$_{c}$=65(5) K), 0.06 and 0.09 (T$_{c}$%
\mbox{$<$}%
2 K) are displayed in Figs. 1e and 1f. The non-magnetic Zn impurities have a
profoundly different impact on the PG phenomenon. The PG onset frequency $%
\omega ^{PG}$ exhibits a slight decrease from 1150 cm$^{-1}$ at y=0 to 890 cm%
$^{-1}$ at y=0.09 (solid squares in\ Fig. 1d) consistent with previous
reports on moderately Zn-substituted Y123 \cite{Uchida1}. In addition, the
PG apparently fills in and becomes less pronounced upon Zn substitution.
This behavior is suggestive of an inhomogeneous scenario where the PG is
locally suppressed around Zn-impurities while it remains virtually
unaffected in remote regions. It agrees with scanning tunneling microscopy
(STM) data on lightly Zn substituted Bi2212 where the spectral gap was
locally suppressed in the vicinity of the Zn impurities \cite{Pan1}. In
contrast, for the Ni impurities there is no indication for such a spatially
inhomogeniety. Especially at high Ni content our spectra even suggest that
the PG develops towards a real gap since $\sigma _{1c}^{el}$ becomes almost
fully suppressed at nonzero frequency. The continuous increase of $\omega
^{PG}$ indeed suggests that the Ni impurities interact with the PG
correlations over a sizeable length scale well beyond the nearest neighbor
distance. A corresponding increase of the energy gap was not observed in
previous STM experiments on very lightly Ni substituted samples \cite%
{Hudson1}. We hope that our data will stimulate STM or ARPES experiments on
heavily Ni-substituted crystals.

Our observations provide important information regarding the relevance of
several PG models. Firstly, the apparent Ni impurity-induced enhancement of
the PG energy scale is at variance with proposals that a precursor SC state
lacking macroscopic phase coherence is at the heart of the PG phenomenon %
\cite{Emery1}. The Ni impurities (similar to Zn) give rise to sizeable
potential scattering; the magnetic scattering is considerably weaker and
plays only a minor role in the T$_{c}$ suppression \cite%
{Williams1,Itoh02,Hudson1,Xiang01}. The potential scattering mixes states
with opposite sign of the d-wave SC order parameter and thus gives rise to
destructive interference resulting in a strong T$_{c}$ suppression and also
in a reduction of the magnitude of the energy gap. While a strong-coupling
scenario may explain a fairly weak impurity-induced PG suppression as
observed for the Zn impurities \cite{Levin02}, it does not account for the
observed drastic increase of the PG energy scale upon Ni substitution.
Similar arguments apply for models that account for the PG in terms of
competing order parameters with exotic symmetry like the flux phase or
d-density wave states, which also exhibit a sign change in k-space or
real-space \cite{Lee1,Chakravarty1}. Potential scattering from the Ni
impurities should be deleterious for these order parameters. It is
furthermore difficult to explain how the energy scale of conventional charge
or spin density waves states \cite{Chubukov1,Dahm1}, which rely on the
nesting condition at the Fermi-surface, can be so strongly enhanced by Ni
impurities which broaden the electronic states and thus tend to weaken the
nesting.

While our surprising results allow us to rule out several influential PG
models, it remains an open question how the magnetic Ni impurities give rise
to such a dramatic increase of the PG energy scale. The stunning difference
between magnetic Ni impurities and non-magnetic Zn-impurities points towards
an important role of magnetic correlations. Non-magnetic Zn$^{2+}$ (S=0)
impurities are known to disrupt the network of AF exchange bonds. While low
Zn concentrations (%
\mbox{$<$}
2 \%) may induce magnetic correlations on nearest or next nearest neighbor
Cu sites, these vanish rather rapidly at higher Zn content \cite{Adachi1}.
Our unpublished $\mu $SR experiment establish that the magnetic Ni$^{2+}$
impurities (S=1) progressively strengthen the Cu spin correlations giving
rise to a spin-freezing transition even in optimally and slightly overdoped
samples. This hints at a prominent role of short-range AF spin correlations
in the pseudogap formation. Further neutron scattering studies of the
magnetic excitations in highly Ni-substituted samples \cite{Sidis1}, in
particular, in the relevant energy range of 50-100 meV might allow one to
further clarify the relationship with the PG phenomenon.

A related question is whether the Ni-impurities on the Cu(2) planar or Cu(1)
chains sites are responsible for the PG enhancement. We tried to resolve
this issue by investigating a pair of crystals with y=0.06 where the
relative Ni occupation was modified by a high temperature annealing
treatment (940$%
{{}^\circ}%
$ C in O$_{2}$ as compared to 920$%
{{}^\circ}%
$ C in 0.2 \% O$_{2}$) prior to annealing at 520$%
{{}^\circ}%
$ C. The former sample should contain an appreciably higher Ni-content on
the chains. However, we could not observe any significant shift in $\omega
^{PG}$ which indicates either that we did not manage to induce a sizeable
redistribution of Ni between plane and chain sites, or else that magnetic
impurities on plane and chain sites tend to enhance the PG correlations.
Note that the final answer to this question will not affect our main
conclusions concerning the validity of the various PG models, which rely
solely on the well established fact that Ni impurities on Cu(2) sites give
rise to considerable potential scattering and thus should suppress rather
than enhance a PG order parameter with an exotic symmetry.

Another interesting issue concerns the doping dependence of the PG
phenomenon in the Ni substituted samples. In the pure samples the normal
state PG vanishes at optimum doping, while for overdoped samples the energy
gap formation coincides with T$_{c}$ \cite{Homes1,Bernhard1}. The SC energy
gap thus obscures the evolution of the PG. In particular, it does not allow
one to explore whether the PG merges with the SC one extending well into the
overdoped regime, or whether it vanishes shortly above optimum doping as was
suggested based on specific heat and NMR data \cite{Tallon1}. In the
following we show that our optical data support the latter scenario.
Displayed in Fig. 2 are the c-axis optical spectra for optimally doped but
non-superconducting SmBa$_{2}$Cu$_{2.86}$Ni$_{0.14}$O$_{7}$ (p$\approx 0.16)$
and overdoped Sm$_{0.86}$Ca$_{0.14}$Ba$_{2}$Cu$_{2.88}$Ni$_{0.12}$O$_{7}$
with p$\approx 0.2$ (from TEP). Figure 2a shows that Ni substitution
restores a pronounced PG with $\omega ^{PG}\approx $2200 cm$^{-1}$ in the
optimally doped crystal. For the overdoped crystal, as shown in Fig. 2b,
there remains only a trace of a PG whose energy scale furthermore is
strongly reduced to $\omega ^{PG}\approx $700 cm$^{-1}$. The rapid decline
of the PG energy scale on the overdoped side lends support to models which
relate the onset of the PG correlations to a critical point that is located
well within the SC dome \cite{Tallon1} (and thus hidden in pure samples).

In summary, we have investigated how the normal-state pseudogap (PG) in the
c-axis optical conductivity is affected by magnetic Ni and non-magnetic Zn
impurities. Most surprisingly, we find that the Ni impurities strongly
increase the energy scale of the PG and restore a PG even in optimally and
slightly overdoped samples. Our observations are at variance with theories
that relate the PG to precursor superconductivity where the PG should not be
enhanced by impurities that lead to potential scattering. They also conflict
with models of competing phases with exotic order parameters that are
susceptible to destructive interference effects from potential scattering,
such as flux phase or d-density wave states. Our results indicate that
magnetic correlations play an important role in the PG phenomenon. In
particular, our data highlight that substitution of magnetic Ni impurities
provides a unique opportunity to investigate the ground state properties of
the PG in the absence of superconductivitiy.

We acknowledge the support of Y.L. Mathis and B. Gasharova at ANKA. We thank
M. Mostovoy and R. Zeyher for stimulating discussions. Research at MPI-FKF
was supported by the Deutsche Forschungsgemeinschaft (DFG), grant BE2684/1-2
in FOR 578.

Figure 1: (a)-(c) Real part of the c-axis infrared conductivity
conductivity, $\sigma _{1ab},$ of pure and Ni-substituted underdoped NdBa$%
_{2}\{$Cu$_{1-y}$Ni$_{y}\}_{3}$O$_{6.8}$ single crystals between 10 and 300
K. Arrows mark the onset of the pseudogap at $\omega ^{PG}$. Corresponding
spectra for Zn substituted crystals are shown in (d) and (e). The evolution
of $\omega ^{PG}$ as a function of Ni content (solid diamonds) and Zn
content (open circles) is displayed in (f).

Figure 2: Real part of the c-axis infrared conductivity conductivity, $%
\sigma _{1ab},$ of (a) optimally doped SmBa$_{2}$Cu$_{2.6}$Ni$_{0.4}$O$_{7}$
and (b) overdoped Sm$_{0.9}$Ca$_{0.1}$Ba$_{2}$Cu$_{2.64}$Ni$_{0.36}$O$_{7}$.
(c) The doping dependence of the PG energy scale $\omega ^{PG}$ is shown by
solid squares (circles) for pure underdoped (overdoped) crystals and solid
diamonds for heavily Ni substituted crystals.


\begin{references}
\bibitem{Alloul1} W. W. Warren et al., Phys. Rev. Lett. {\bf 62}, 1193
(1989); H. Alloul et al., Phys. Rev. Lett. {\bf 63}, 1700 (1989).

\bibitem{Loram1} J.W. Loram et al., Phys. Rev. Lett. {\bf 71}, 1740 (1993).

\bibitem{Loeser1} A.G. Loeser et al., Science {\bf 273}, 325 (1996).

\bibitem{Homes1} C.C. Homes et al., Phys. Rev. Lett. {\bf 71}, 1645 (1993).

\bibitem{Bernhard1} C. Bernhard et al., Phys. Rev. {\bf B 59}, 6631 (1999)
and unpublished data.

\bibitem{Emery1} V.J. Emery and S.A. Kivelson, Nature {\bf 374}, 434 (1995).

\bibitem{Alexandrov1} A.S. Alexandrov, V.V. Kabanov, and N.F. Mott, Phys.
Rev. Lett. {\bf 77}, 4796 (1996).

\bibitem{Tallon1} for a review see, J.L. Tallon et al., Phys. Stat. Sol. 
{\bf (b) 215}, 531 (1999).

\bibitem{Chubukov1} A.V. Chubukov, and J. Schmalian, Phys. Rev. {\bf B 57},
11085 (1998).

\bibitem{Dahm1} T. Dahm, D. Manske, and L. Tewordt, Phys. Rev. {\bf B 56},
11419 (1997).

\bibitem{Lee1} P.A. Lee, and X.G. Wen, Phys. Rev. Lett. {\bf 76}, 503 (1996).

\bibitem{Chakravarty1} S. Chakravarty et al., Phys. Rev. {\bf B 64}, 094503
(2001).

\bibitem{Loew1} U. L\"{o}w et al., Phys. Rev. Lett. {\bf 72}, 1918 (1994).

\bibitem{Anderson1} P.W. Anderson, Science {\bf 235}, 1196 (1987).

\bibitem{DiCastro1} C. Castellani, C. DiCastro, and M. Grilli, Phys. Rev.
Lett. {\bf 75}, 4650 (1995).

\bibitem{Uchida1} Y. Fukuzumi, K. Mizuhashi, and S. Uchida, Phys. Rev. {\bf %
B 61}, 627 (2000); R. Hauff et al., Phys. Rev. Lett. {\bf 77}, 4620 (1996).

\bibitem{WolfRE} Th. Wolf et al., J. Crystal Growth {\bf 96}, 1010 (1989).

\bibitem{TallonZnPRL} J. L. Tallon et al., Phys. Rev. Lett. {\bf 79}, 5294
(1997).

\bibitem{Williams1} G.V.M. Williams, R. Dupree, and J.L. Tallon, Phys. Rev. 
{\bf B 60}, 1360 (1999); Phys. Rev. {\bf B 61}, 4319 (2000).

\bibitem{Itoh02} Y. Itoh, S. Adachi et al., J. Bobroff et al., Phys. Rev.
Lett. {\bf 79}, 2117 (1997).

\bibitem{TallonTEP} J. L. Tallon et al., Phys. Rev. Lett. {\bf 75}, 4114
(1995); Phys. Rev. {\bf B 51}, 12911 (1995.

\bibitem{Bernhard2} C. Bernhard, J. Humlicek and B. Keimer, Thin Solid Films 
{\bf 455-456}, 143 (2004).

\bibitem{Wolfnh} S.I. Schlachter et al., Int. J. Modern Phys. {\bf B 14},
3673 (2000).

\bibitem{Grueninger1} M. Gr\"{u}ninger et al., Phys. Rev. Lett. {\bf 84},
1575 (2000); D. Munzar et al., Solid State Commun. {\bf 112}, 365 (1999).

\bibitem{Pan1} S.H. Pan et al., Nature {\bf 403}, 746 (2000).

\bibitem{Hudson1} E.W. Hudson et al., Nature {\bf 411}, 920 (2001).

\bibitem{Xiang01} T. Xiang et al., Phys. Rev. {\bf B 66}, 174504 (2002).

\bibitem{Levin02} Y.-J. Kao et al., Phys. Rev. {\bf B 66}, 214519 (2002).

\bibitem{Adachi1} T. Adachi et al., Phys. Rev. {\bf B 69}, 184507 (2004).

\bibitem{Sidis1} Y. Sidis et al., Phys. Rev. Lett. {\bf 84}, 5900 (2000).

\bigskip \newpage
\end{references}
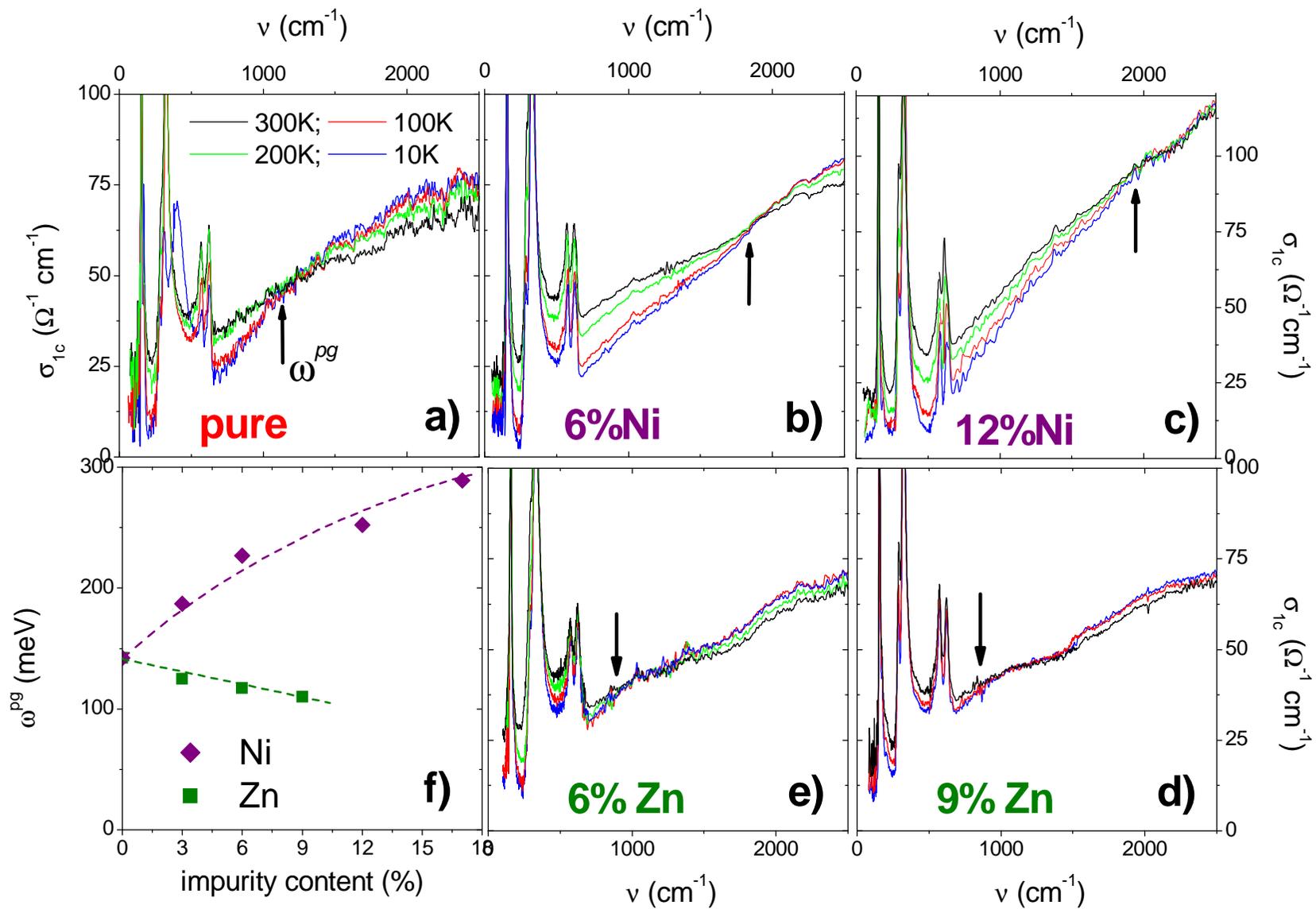
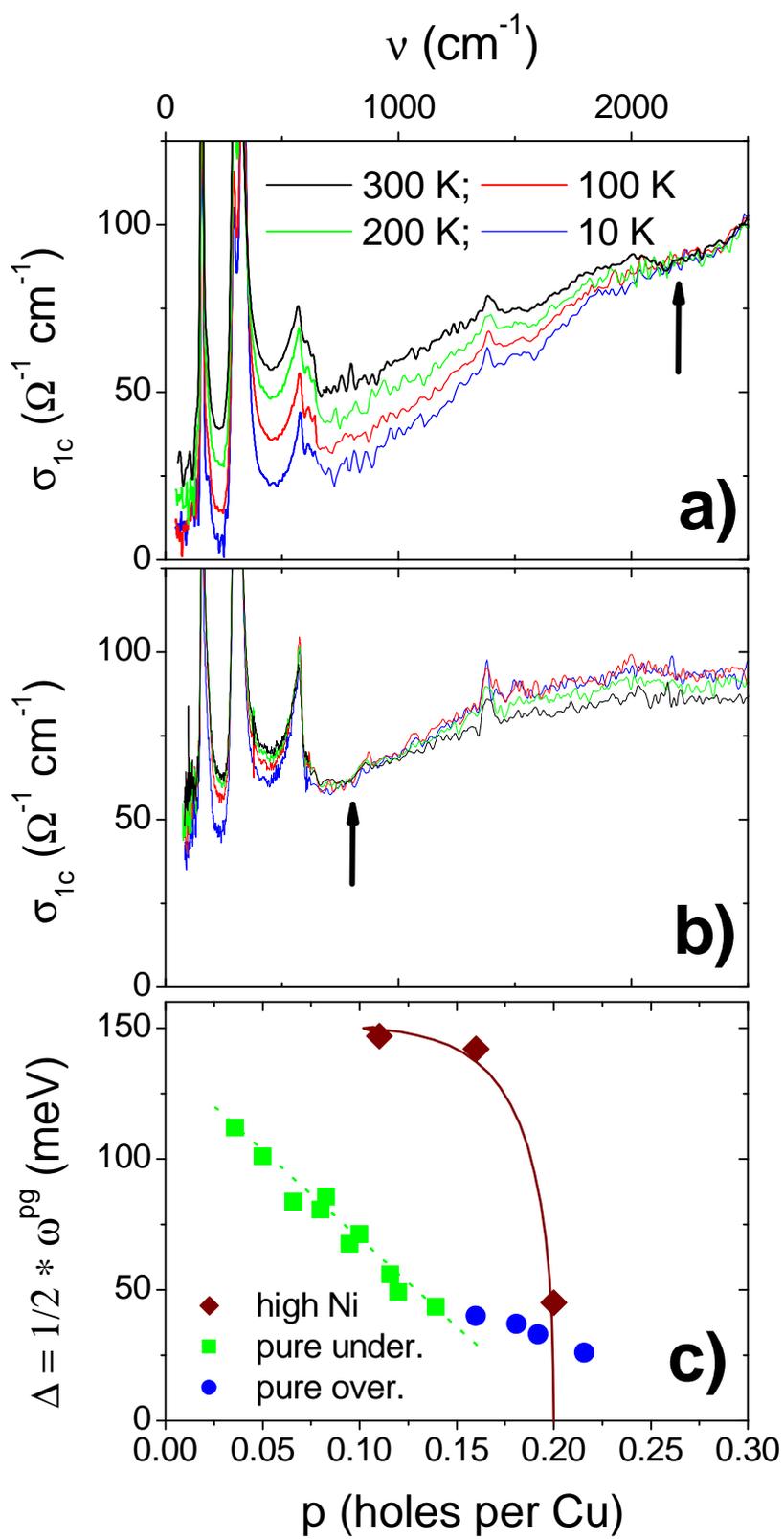
\end{document}